\documentclass[%
 prl,
 amsmath,amssymb,
 reprint,%
 superscriptaddress,
]{revtex4-1}
\usepackage{graphicx}
\usepackage{dcolumn}
\usepackage{bm}
\usepackage{color,graphicx}
\usepackage{amsmath}
\usepackage{eqnarray}
\usepackage{amssymb}
\usepackage{amsthm}
\usepackage{amsfonts}
\usepackage{braket}
\usepackage[arrowdel]{physics}
\pdfoutput=1 
\begin{document}

\title{Unified Model for Probing Solar Cell Dynamics via Cyclic Voltammetry and Impedance Spectroscopy}

\author{Victor Lopez-Richard}
\affiliation{Departamento de Física, Universidade Federal de São Carlos, 13565-905 São Carlos, SP, Brazil}
\email{vlopez@df.ufscar.br}

\author{Luiz A. Meneghetti Jr }
\affiliation{Departamento de Física, Universidade Federal de São Carlos, 13565-905 São Carlos, SP, Brazil}

\author{Gabriel L. Nogueira}
\affiliation{São Paulo State University (UNESP), School of Sciences, Department of Physics and Meteorology, Bauru, SP, 17033-360, Brazil}

\author{Fabian Hartmann}
\affiliation{Julius-Maximilians-Universität Würzburg, Physikalisches Institut and Würzburg-Dresden Cluster of Excellence ct.qmat, Lehrstuhl für Technische Physik, Am Hubland, 97074 Würzburg, Deutschland}

\author{Carlos F. O. Graeff }
\affiliation{São Paulo State University (UNESP), School of Sciences, Department of Physics and Meteorology, Bauru, SP, 17033-360, Brazil}

\date{\today}

\begin{abstract}
Despite the remarkable progress in emerging solar cell technologies such as hybrid organic-inorganic perovskites, there are still significant limitations related to the stability of the devices and their non-ideal electrical behavior under certain external stimuli. We present a conceptual framework for characterizing photovoltaic devices by integrating cyclic voltammetry (CV) and impedance spectroscopy (IS). This framework is constructed from a microscopic, multi-mode perspective that explicitly accounts for drift, diffusion, displacement, and memory contributions. We derive comprehensive analytical expressions for current-voltage relationships and complex admittance. Our model reveals the inseparable connection between hysteresis behaviors in current-voltage characteristics observed in CV and the apparent capacitive and inductive behaviors seen in IS spectral analysis. We demonstrate how CV and IS naturally complement each other, providing a deeper microscopic understanding of device performance and limitations. Additionally, we establish the relationship between intrinsic material parameters and experimentally accessible extrinsic parameters such as light intensity, temperature, DC bias, voltage amplitude, and frequency. This framework enables unprecedented optimization of solar cell performance, marking a significant advancement towards sustainability.
\end{abstract}

                              
\maketitle


Solar cells based on novel materials and material combinations such as perovskites~\cite{Green2014}, organics~\cite{Sun2022}, inorganic-organic hybrids~\cite{Aqoma2024}, and quantum dots~\cite{Ning2015} have attracted  attention in recent years and opened new avenues towards a sustainable, energy-aware society. While solar cells based on silicon lead the market due to their state-of-the-art low-cost fabrication and reliability, they are ultimately limited in terms of device performance. In contrast, solar cells based on emergent materials offer unprecedented advantages compared to silicon solar cells, such as the realization of solar cells on flexible substrates, higher power conversion efficiencies, and the potential for lower manufacturing costs. For instance, perovskite solar cells allow for the fabrication of lightweight and versatile photovoltaic devices~\cite{Kang2019,Wang2018}. While these devices show excellent in-lab efficiencies, they suffer from poor reliability, reproducibility, and instabilities~\cite{Thiesbrummel2024,Chowdhury2023,Baumann2023,Zhang2024,Zhang2024b,Ren2024}. To address these challenges, strategies are actively developed to improve material stability, optimize fabrication processes, and enhance the overall performance of these emerging solar cell technologies~\cite{Sidhik2024}.

In this context, understanding the interplay between electrochemical and electronic processes within solar cells is crucial for optimizing their performance. Cyclic voltammetry (CV)~\cite{Kavan2020,Alvarez2020,Clarke2023,Lee2017} and impedance spectroscopy (IS)~\cite{Chae2016,Zolfaghari2019,Ebadi2019,Dualeh2014} offer a powerful combined approach for achieving this goal. CV unveils the secrets of charge transfer, revealing recombination processes, intrinsic symmetry constraints, and energy levels of active materials~\cite{Thiesbrummel2024}. IS dissects the device's electrical landscape, providing a detailed map of charge transport dynamics~\cite{Ito2008,Bredar2020}, apparent resistance, apparent capacitance, and apparent inductance operating at different time scales~\cite{MoraSero2006,Proskuryakov2007,Guerrero2016}. These combined techniques are instrumental in the quest to improve device efficiency, stability, and overall performance. However, interpreting the data isn't always straightforward. The complexity of overlapping processes and the need for accurate models to disentangle these phenomena pose significant challenges.

Within the realm of dynamics effects in solar cells, two key areas of debate emerge: hysteresis with diverse shapes in current-voltage (I-V) characteristics~\cite{Wu2018,Alvarez2020} (ascribed to memory related processes) and the interplay between capacitive and inductive-like transport responses~\cite{Ebadi2019,Gonzales2022}. Numerous approaches have been proposed to address these complexities~\cite{Takahashi2019,Bou2020,Bisquert2023}. However, limitations remain, including simplifying assumptions, the need for robust hypothesis validation, and the potential for alternative explanations. These are currently hot topics in the field, and researchers are actively seeking to develop more comprehensive models that can fully capture these complex phenomena~\cite{Filipoiu2022,vanNijen2023,Bisquert2024}.

Our model is designed to encompass fundamental principles of photovoltaic systems, incorporating both diffusion and drift contributions under non-equilibrium conditions. It offers a broad applicability without being confined to specific architectures or materials. The model can be directly applied to basic p-n junctions, p-i-n structures with thin absorbers, and p-i-n architectures under strong absorption conditions where recombination in the absorption layer is negligible~\cite{Crandall1983}.

To gain a deeper understanding of how solar cells respond to cyclic voltage inputs, common in CV and IS techniques, we propose a comprehensive approach that builds upon well established physical principles. We have undertaken the challenge of developing a minimalistic, unified theoretical framework capable of simulating both the CV response and the IS analysis. 
This dynamic characterization enables the assessment of the character of fill factor losses due to carrier leakage through the qualitative analysis of defect levels and their activation time scales, as well as evaluation of carrier diffusivity efficiency and the impact of unavoidable geometric capacitances.
Thus, our model aspires to capture the essential physical processes governing the device's behavior under both static and dynamic operating conditions, offering also a framework for the analysis of the apparent controversial issues highlighted above. This approach should serve as a springboard for subsequent discussions and as reference ground for adding complexity layers.
To maintain a qualitative perspective in our discussion, we have omitted units in the figures; however, all units can be readily extracted from the expressions provided in the text.

\begin{figure}[t]
	\includegraphics[width=8.5cm]{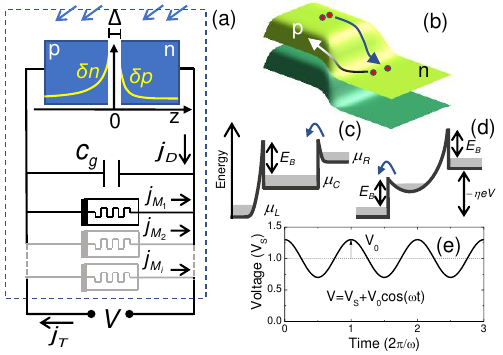}
	\caption{\label{Figure1} (a) Equivalent circuit configuration used for the simulation of the solar cell response. (b) 3D band profile of the junction highlighting the presence of memory channels with memristive nature. Band profile representations of a trapping and a generation site in (c) and (d), respectively. (e) Voltage input used for the solar cell characterization during both cyclic voltammetry and impedance analysis. }
\end{figure}

For an arbitrary bias voltage, $V$, the photo-diode response (as the general category of devices that includes solar cells) can be decomposed in three independent contributions, as represented in Fig.~\ref{Figure1} (a),
\begin{equation}
    j_T=j_D+\frac{C_g}{A} \frac{dV}{dt}+\sum_i j_{M_i},
    \label{jt}
\end{equation}
where
\begin{equation}
j_D=e D_n \frac{\partial n}{\partial z}|_{z= -\frac{\Delta}{2}}-e D_p \frac{\partial p}{\partial z}|_{z= \frac{\Delta}{2}}, 
    \label{jd}
\end{equation}
is the diode current density obtained by adding the minority carriers diffusive components at the boundaries of the depletion (or intrinsic) region of width $\Delta$ (with axis origin at the midpoint), where the approximation of uniform electron and hole current components is assumed~\cite{Green1982}. The second term in Eq.~\ref{jt} accounts for the displacement contribution given the geometric capacitance, $C_g$, of the device and its area, $A$. The third term combines all the ionic channels, fluctuations of drift components of generated carriers at the junction (or intrinsic region), and even potential leakage pathways, as,
\begin{equation}
j_{M_i}=\gamma_i (N_i^0+\delta N_i)V,
   \label{jm}
\end{equation}
for $N_i^0+\delta N_i$ carriers that contribute to the conductance, with $\gamma_i=\frac{e \mu_i}{A\Delta^2}$, and mobility $\mu_i$ along the length $\Delta$.
We assume each contribution to be an independent fluctuation of non-equilibrium carriers around $N_i^0$ described by certain relaxation time, $\tau_i$~\cite{Silva2022,Paiva2022},
\begin{equation}
    \frac{d \delta N_i}{d t}=-\frac{\delta N_i}{\tau_i}+g_i(V).
    \label{dn}
\end{equation}
The combination of the ingredients in Eqs.~\ref{jm} and~\ref{dn} has been demonstrated to be sufficient to consistently induce memory responses~\cite{Silva2022,Paiva2022,LopezRichard2022}. Therefore, we will refer to the contributions of the terms in Eqs.~\ref{jm} as the memory components.

A representation of this process has been provided in Fig.~\ref{Figure1} (b) as carrier leakage through the surface, though it is not necessarily limited to superficial effects since crystal defects and impurity induced precipitates in the junction (or intrinsic) region can also play a role~\cite{Green1982}. 
The dynamics of non-equilibrium charges described by Eq.~\ref{dn} encompass contributions from both extrinsic and photo-generated carriers, as well as any ionic motion~\cite{Ravishankar2017,Ghahremanirad2017} influenced by the built-in electric field.

For simplicity, we will restrict the discussion to a single memory channel and drop the subindex $i$ in what follows. These carriers can be trapped or released by thermal ionization at a temperature 
$T$, according to the localization profiles displayed in Figs.~\ref{Figure1} (c) and (d) which encompass a wide range of possibilities. According to Ref.~\citenum{LopezRichard2022}, the carrier generation or trapping rate is described in these cases by
\begin{equation}
    g(V)=\frac{i_0}{\eta}\left[e^{-\eta_L \frac{eV}{k_B T}}+e^{\eta_R \frac{eV}{k_B T}} -2 \right],
    \label{gen}
\end{equation}
where $i_0=\frac{4 \pi m^* A}{(2 \pi \hbar)^3} (k_B T)^2  e^{-\frac{E_B}{k_B T}}$, $\eta_L=\alpha \eta/(1+\alpha)$, and $\eta_R= \eta/(1+\alpha)$. Here, $\alpha \in [0,\infty)$ characterizes the symmetry of the carrier transfer in Eq.~\ref{gen} with respect to the local bias voltage drop, where $\alpha=1$ corresponds to the symmetric case. Note that in the limits as $\alpha \rightarrow 0$ or $\alpha \rightarrow \infty$, the function $g(V)$ becomes almost insensitive to either negative or positive large bias, respectively, as the first or second term in brackets tend to $1$.
Additionally, $\eta<0$ corresponds to the diagram in Fig.~\ref{Figure1} (c), while $\eta>0$ corresponds to the diagram in Fig.~\ref{Figure1} (d) with $1/|\eta|$ being the number of localization sites along a single line of the device length. Thus it is reasonable to expect that $|\eta|<<1$. A range of carrier generation and trapping processes can be effectively captured by expressions similar to Eq.~\ref{gen}, which represent activation fluxes over barriers with heights modulated by voltage~\cite{Ravishankar2017,Ghahremanirad2017}. These barriers may arise at interfaces and surfaces, local defects, at crystallite or grain boundaries in polycrystalline or amorphous materials. A detailed discussion on how these microscopic parameters, such as barrier architectures and symmetry constraints, modulate carrier transfer is provided in Ref.~\citenum{LopezRichard2022}.

While triangular voltage sweeps are the traditional choice for CV characterization, we propose that sinusoidal sweeps offer a more natural connection to IS. Since IS utilizes harmonic biases, employing sinusoidal voltages in CV aligns these techniques both conceptually and potentially from a modeling and interpretation standpoint. This unification facilitates the development of the single framework for discussion as presented below enabling a seamless correlation between CV and IS data.

Thus, we will assume a general case of a combination of a stationary DC voltage and an alternating AC voltage, expressed as $V=V_S+V_0 \cos(\omega t)$ and represented in Fig.~\ref{Figure1} (e). This setup encompasses the conditions for CV when $V_S=0$ and impedance spectroscopy characterization for arbitrary $V_S$. Next, we propose to extend the method presented in Ref.~\citenum{Pikus1965} for a simple p-n junction to the case of a solar cell using a multimode expansion that incorporates the ineluctable generation of higher harmonics. This approach enables us to solve Eq.~\ref{jd} under illumination and cyclic voltammetry conditions with an arbitrary AC amplitude, $V_0$. 

We should start from the continuity equations and the Ficks law, with diffusion coefficients $D_p$ and $D_n$ for holes and electrons, respectively, assuming a uniform optical electron-hole generation in the volume, $g_L$, \cite{Green1982}
\begin{eqnarray}
    \frac{d \delta p}{d t}&=&-\frac{\delta p}{\tau_p}+g_{L}+D_p \frac{d^2 \delta p}{d z^2}, \nonumber \\
    \frac{d \delta n}{d t}&=&-\frac{\delta n}{\tau_n}+g_{L}+D_n \frac{d^2 \delta n}{d z^2}.
    \label{cont}
\end{eqnarray}
Here, the conventional boundary conditions for the minority carrier fluctuation are given by: $\delta p \left(\frac{\Delta}{2} \right)=p_{eq} \left( e^{\frac{eV}{k_B T}} - 1\right)$,  $\delta p(\infty)=0$, $\delta n \left(-\frac{\Delta}{2} \right)=n_{eq} \left( e^{\frac{eV}{k_B T}} - 1\right)$,  $\delta n(-\infty)=0$, as represented in Fig.~\ref{Figure1} (a), where, $p_{eq}$ and $n_{eq}$ are the minority carriers equilibrium densities as provided by the mass action law. Note that, as a general case, we may consider contrasting recombination times $\tau_p$ and $\tau_n$ for electrons and holes, respectively. Using the generalization of the voltage input as complex function, $V=V_S+V_0 e^{i \omega t}$, the stable solutions of Eqs.~\ref{cont} can be sought in the most general form as (once any transient processes have decayed)
\begin{eqnarray}
    \delta p (z)=\sum_{m=-\infty}^{\infty} P_m(z) e^{i m \omega t}, \nonumber \\
    \delta n (z)=\sum_{m=-\infty}^{\infty} N_m(z) e^{i m \omega t},
    \label{pn}
\end{eqnarray}
which corresponds to an unavoidable multimode perspective of the dynamic response, as explored in what follows. By substituting Eqs.~\ref{pn} into Eqs.~\ref{cont} we obtain, for holes
\begin{equation}
    \sum_{m=-\infty}^{\infty}\left(D_p \frac{d^2 P_m}{d z^2} +g_L \cdot \delta_{m,0} -\frac{P_m}{\tau_p} + i m \omega P_m \right)e^{i m \omega t}=0,
    \label{holes1}
\end{equation}
with $\delta_{m,n}$ representing the Kronecker delta. Thus, the solutions for $P_m$ can be readily calculated as
\begin{eqnarray}
    P_m(z)&=&\left[P_m\left(\frac{\Delta}{2} \right)-g_L \tau_p\cdot \delta_{m,0}  \right] e^{-(z-\frac{\Delta}{2})/L_p^{(m)}} \nonumber \\
    &+& g_L \tau_p \cdot \delta_{m,0},
    \label{pm}
\end{eqnarray}
with 
\begin{equation}
L_p^{(m)}=\frac{L_p}{\sqrt{1+ i m \omega \tau_p} },
   \label{lm}
\end{equation}
where $L_p=\sqrt{D_p \tau_p}$ is the diffusion length. The resulting equation for electrons is analogous, by replacing the subindex $p \rightarrow n$, $z \rightarrow -z$, and $\Delta \rightarrow -\Delta$. 

We may now combine the results in Eqs.~\ref{pm} and~\ref{pn} into Eq.~\ref{jd}, yielding 
\begin{equation}
j_D=\sum_{m=-\infty}^{\infty} \left[ e D_n \frac{N_m\left(-\frac{\Delta}{2} \right)}{L_n^{(m)}} + e D_p \frac{P_m\left(\frac{\Delta}{2} \right)}{L_p^{(m)}} \right] e^{i m \omega t}.
   \label{jdt}
\end{equation}
Then, by using the boundary conditions, $\delta p\left(\frac{\Delta}{2} \right)$ and $\delta n\left(-\frac{\Delta}{2} \right)$, the coefficients $P_m\left(\frac{\Delta}{2} \right)$ and $N_m\left(-\frac{\Delta}{2} \right)$ can be easily obtained noting that, according to Eqs.~\ref{pn}
\begin{eqnarray}
p_{eq} \left( e^{\frac{eV}{k_B T}} - 1\right)&=&\sum_{m=-\infty}^{\infty} P_m\left(\frac{\Delta}{2} \right) e^{i m \omega t}, \nonumber \\
n_{eq} \left( e^{\frac{eV}{k_B T}} - 1\right)&=&\sum_{m=-\infty}^{\infty} N_m\left(-\frac{\Delta}{2} \right) e^{i m \omega t},
\end{eqnarray}
and that $e^{\frac{eV}{k_B T}}=e^{\frac{eV_S}{k_B T}}e^{\frac{eV_0 \exp(i \omega t)}{k_B T}}$ can be expanded as an infinite Taylor series 
\begin{equation}
e^{\frac{eV}{k_B T}}=e^{\frac{eV_S}{k_B T}} \sum_{m=0}^{\infty}  \frac{1}{m!}\left(\frac{eV_0 }{k_B T}\right)^m e^{i m \omega t},
   \label{exp}
\end{equation}
with no restrictions to the value of $\frac{eV_0 }{k_B T}$, yielding
\begin{widetext}
\begin{eqnarray}
j_D&=& \left( \frac{eD_p p_{eq}}{L_p} + \frac{eD_n n_{eq}}{L_n}\right)\left( e^{\frac{e V_S}{k_B T}} - 1\right) - e g_L \left( L_n+L_p \right) \nonumber \\
&+&\frac{e^{\frac{e V_S}{k_B T}}}{\sqrt{2}}\sum_{m=1}^{\infty}  \left\{ \frac{e D_p p_{eq}}{L_p}\sqrt{1+\sqrt{1+(m\omega \tau_p)^2}} + \frac{e D_n n_{eq}}{L_n}\sqrt{1+\sqrt{1+(m\omega \tau_n)^2}}\right. \nonumber \\
&+&i \left. \left[ \frac{e D_p p_{eq}}{L_p}\sqrt{\sqrt{1+(m\omega \tau_p)^2}-1} + \frac{e D_n n_{eq}}{L_n}\sqrt{\sqrt{1+(m\omega \tau_n)^2}-1} \right]   \right\} \frac{1}{m!}\left(\frac{eV_0 }{k_B T}\right)^m e^{i m \omega t}.
   \label{jdfin}
\end{eqnarray}
\end{widetext}
Here, the identity $\sqrt{1+i a}=1/\sqrt{2}(\sqrt{1+\sqrt{1+a^2}}+i\sqrt{\sqrt{1+a^2}-1} )$ for $a>0$ has been used. Note that in Eq.~\ref{jdfin} the diffusive channels contribute to both resistive (real part contributions) and reactive terms (imaginary part contributions) that cannot be reduced to simple elementary circuit components without attributing to them a complex frequency dependence. It is important to point out however that the contribution to the susceptance (imaginary part) is positive for all modes, indicating a capacitive character across the board of the diffusive terms.    

The case of the memristive contribution, $j_M$, is simpler to handle analytically. Although the condition $\frac{e V_0}{k_B T}<<1$ cannot always be assumed for cyclic voltammetry or large amplitude spectroscopy, the condition $|\eta| \frac{e V_0}{k_B T}<<1$, is more achievable due to the typically small absolute value of $\eta$ in Eq.~\ref{gen}. Thus, by expanding the generation function up to second order in $|\eta| \frac{e V_0}{k_B T}$ and solving Eq.~\ref{dn} we can fully understand the topological nuances of the dynamic effect of the memory contributions on the current-voltage response of the solar cell. In this context, the term topology refers to the shapes of the hysteresis loops in the current-voltage response. This solution is an extension of the methodology described in Refs.~\citenum{LopezRichard2022} and~\citenum{LopezRichard2024}. For the stable case, once the transient contributions depending on the initial conditions fade, it can be expressed up to third order mode as
\begin{widetext}
\begin{eqnarray}
j_M&=& \gamma \left( N_0 V_S +g_0 \tau\right) + \frac{N V_S}{2}+ \frac{M V_0}{1+(\omega \tau)^2} \nonumber \\
&+& \left\{ \gamma \left( N_0 + g_L \tau + g_0 \tau \right)+ \frac{2 M V_s/V_0}{1+(\omega \tau)^2}(1- i \omega \tau) + \frac{N}{2} \left[ 1+\frac{1/2}{1+(2\omega \tau)^2}\left( 1- i 2 \omega \tau  \right)   \right]  \right\} V_0 e^{i \omega t} \nonumber \\
&+& \left\{ \frac{N V_s/V_0}{2}\frac{1}{1+(2 \omega \tau)^2}(1- i 2 \omega \tau) 
     +  \frac{M}{1+(\omega \tau)^2}(1- i \omega \tau)  \right\}V_0 e^{i 2 \omega t} \nonumber \\
&+&  \left\{ \frac{N }{4}\frac{1}{1+(2 \omega \tau)^2}(1- i \omega \tau)\right\} V_0 e^{i 3 \omega t}.
   \label{jmfin}
\end{eqnarray}
\end{widetext}
Here,
\begin{equation}
    M=\gamma \frac{i_0}{2}\frac{eV_0}{k_B T}\tau \frac{e^{\eta_R \frac{eV_S}{k_B T}} - \alpha e^{-\eta_L \frac{eV_S}{k_B T}} }{1+ \alpha},
    \label{M}
\end{equation}
\begin{equation}
    N=\gamma \frac{i_0}{2}\left(\frac{eV_0}{k_B T} \right)^2\tau \eta \frac{e^{\eta_R \frac{eV_S}{k_B T}} + \alpha^2 e^{-\eta_L \frac{eV_S}{k_B T}} }{\left( 1+ \alpha \right)^2},
    \label{N}
\end{equation}
and
\begin{equation}
    g_0= \frac{i_0}{\eta} \left( e^{\eta_R \frac{eV_S}{k_B T}} +  e^{-\eta_L \frac{eV_S}{k_B T}} - 2\right).
\end{equation}
The contribution of higher-order terms in $|\eta| \frac{e V_0}{k_B T}$ can subsequently be obtained using the same procedure or by numerically solving Eq.~\ref{dn}. 
\begin{figure}[h]
	\includegraphics[width=8.5cm]{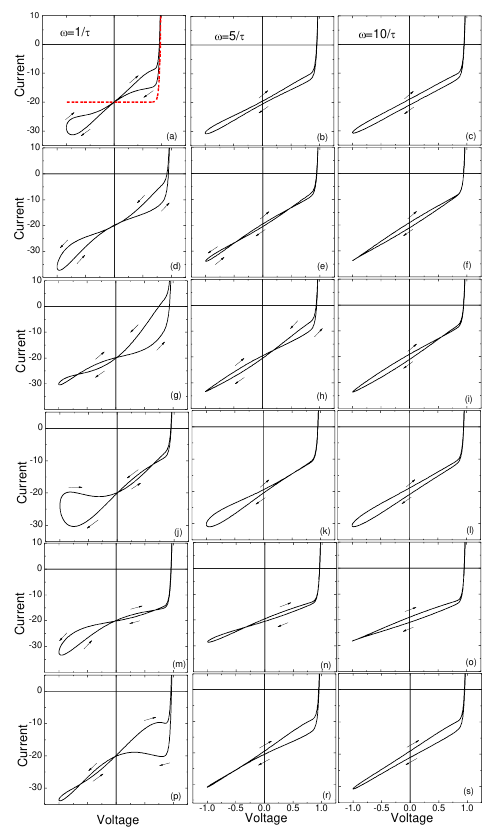}
	\caption{\label{Figure2} Cyclic voltammetry hysteresis of an illuminated solar cell, each column corresponding to a different AC voltage frequency: (a), (b), and (c) $\alpha=1$ and $\eta<0$; (d), (e), and (f) $\alpha=1$ and $\eta>0$; (g), (h), and (i) $\alpha=0.99$ and $\eta>0$; (j), (k), and (l) $\alpha=0.99$ and $\eta<0$; (m), (n), and (o) $\alpha=1.01$ and $\eta>0$; (p), (q), and (r) $\alpha=1.01$ and $\eta<0$. The unperturbed diode current-voltage characteristic has been added to panel (a) as a dashed line for reference.}
\end{figure}
Within this notation, the displacement contribution to Eq.~\ref{jt} is given by $i \frac{C_g}{A} \omega V_0 e^{i \omega t}$. With this, we now have all the ingredients needed to analyze the total solar cell current response, $j_T$, to cyclic voltage inputs of arbitrary amplitudes and frequencies in the framework of the transport models used as starting hypotheses. 

Let's start by describing the potential result of CV, for which $V_S=0$ and the frequency is usually such that $\omega << \min [1/\tau_n, 1/\tau_p]$. Under this conditions, the first term in Eq.~\ref{jdfin} vanishes. To handle the infinite series, we can take advantage of the low-frequency condition and approximate $\omega \tau_n = \omega \tau_p \rightarrow 0$ that also cancels the reactive contribution (imaginary component of the current) in the third line of Eq.~\ref{jdfin}. This allows us to evaluate the infinite sum in Eq.~\ref{jdfin} directly, for arbitrary large voltage amplitude, $V_0$, resulting in the following expression for the real part of the diode current
\begin{equation}
\Re{j_D}= j_S\left( e^{\frac{e V_0 \cos{\omega t}}{k_B T}} - 1\right) - e g_L \left( L_n+L_p \right),
\label{jdcv}
\end{equation}
with $j_S= \left( \frac{eD_p p_{eq}}{L_p} + \frac{eD_n n_{eq}}{L_n}\right)$. The expression in Eq.~\ref{jdcv} corresponds to the dashed reference curve in Fig.~\ref{Figure2}, and can be added to the real parts of the displacement contribution and to Eq.~\ref{jmfin} leading to the results displayed in Fig.~\ref{Figure2} as solid lines for a solar cell under illumination. The idea of the figure is to illustrate a variety of hysteresis loops that can be obtained by tuning two key parameters of the memory channel: the symmetry factor $\alpha$ and the nature of the non-equilibrium carrier transfer, determined by the sign of $\eta$. Panels (a)-(c) depict symmetric trapping, while panels (d)-(e) illustrate symmetric generation processes. In the remaining panels, a certain degree of asymmetry ($\alpha \neq 1$) has been introduced, contributing to the polarity dependence of the CV shape. This asymmetry highlights the nuanced relationship between trapping and generation mechanisms under different voltage conditions.

The shape of the hysteresis is influenced not only by intrinsic non-equilibrium mechanisms but also, critically, by the characteristics of the external drive. Hysteresis patterns should not be considered definitive signatures of the device response without a precise description of the applied voltage pulses. The distinction between 'normal' and 'inverted' hysteresis is inherently linked to the specifics of system excitation, including pulse shapes, amplitudes, and periods. With this clarification, the direction of the hysteresis loops can serve as a valuable tool for characterizing underlying microscopic processes.

Note that, despite the rich variety of responses at frequencies close to the condition $\omega \sim 1/\tau$, all the hysteresis loops converge to clockwise open loops as the frequency increases from the left to the right column of Fig.~\ref{Figure2}. This occurs because all the reactive components in Eq.~\ref{jmfin} vanish when $\omega \tau >> 1$ causing the displacement contribution to the reactive part of the total current density, proportional to $\omega C_g V_0$, to dominate. The interplay and frequency tuning of clockwise and counter-clockwise loops in CV (sometimes lieading to multiple crossings) result from combinations of nonequilibrium carrier trapping and generation, respectively. These behaviors can be described as capacitive or inductive based on the apparent anticipation or delay of the current with respect to the voltage sweep. However, the microscopic origins of these dynamic responses are more effectively characterized using IS methods that offer a more intuitive decomposition of each contributing factor, providing clearer insights into the individual mechanisms driving the observed phenomena.

Equations~\ref{jdfin} and \ref{jmfin} reveal the frequency dependence of the current density, suggesting that a spectral analysis based solely on a single mode will be incomplete. To address this, a multimode perspective is necessary. In this framework, the total current can be decomposed as
\begin{equation}
    I_T(t)=A\cdot j_T(t)= \sum_{m=0}^{\infty} \left[G^{(m)}(\omega) +i B^{(m)}(\omega)  \right] V_0 e^{i m \omega t}.
    \label{IT}
\end{equation}
where $G^{(m)}$ and $B^{(m)}$ represent the conductance and susceptance, respectively, for the m-th mode. This decomposition allows us to define the impedance per mode~\cite{LopezRichard2024} as, $Z^{(m)}=\left[G^{(m)}(\omega) +i B^{(m)}(\omega)  \right]^{-1}$. While this work explores the multimode behavior, acknowledging the prevailing focus on fundamental mode analysis in traditional impedance studies, we will highlight the results for the $m = 1$ mode.

\begin{figure}[h]
	\includegraphics[width=8.5 cm]{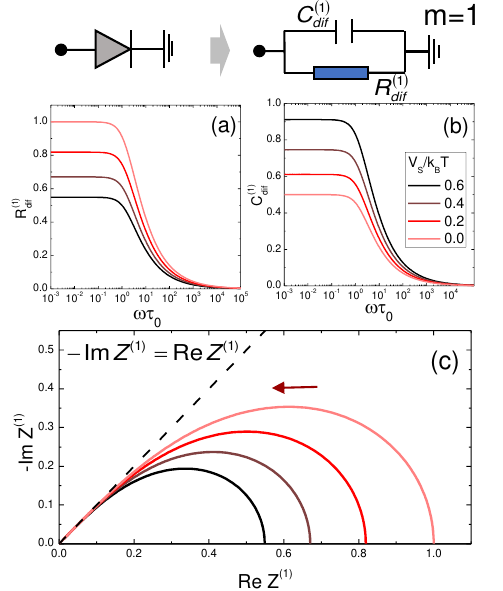}
	\caption{\label{Figure3new} Upper panel: Schematic representation of the first-order mode of the diode-like drift-diffusion component. (a) Apparent diffusive resistance as a function of voltage frequency for various DC bias values. (b) Corresponding apparent diffusive capacitance. (c) Nyquist plot of the impedance for the first-order mode of the diode current, with the arrow indicating the direction of increasing frequency. The dashed line represents the asymptotic limit at high frequencies, where $-\Im Z^{(1)}=\Re Z^{(1)}$. The assumption $\tau_p=\tau_n=\tau_0$ has been applied throughout  all panels.}
\end{figure}

Within this perspective, the $m=1$ impedance contribution of the diode component in Eq.~\ref{jdfin}, allows defining an apparent diffusive resistance, as represented in the upper panel of Fig.~\ref{Figure3new},
\begin{widetext}
\begin{equation}
 \frac{1}{R_{dif}^{(1)}(\omega)} \equiv \Re \frac{1}{Z^{(1)}}= \frac{e A e^{\frac{e V_S}{k_B T}}}{k_B T\sqrt{2}}\left[ \frac{e D_p p_{eq}}{L_p}\sqrt{\sqrt{1+(\omega \tau_p)^2}+1} + \frac{e D_n n_{eq}}{L_n}\sqrt{\sqrt{1+(\omega \tau_n)^2}+1} \right], 
   \label{RD}
\end{equation}
and an effective diffusive capacitance
\begin{equation}
C_{dif}^{(1)}(\omega)\equiv \Im \frac{1}{Z^{(1)}} \omega^{-1}= \frac{e A e^{\frac{e V_S}{k_B T}}}{\omega k_B T\sqrt{2}}\left[ \frac{e D_p p_{eq}}{L_p}\sqrt{\sqrt{1+(m\omega \tau_p)^2}-1} + \frac{e D_n n_{eq}}{L_n}\sqrt{\sqrt{1+(m\omega \tau_n)^2}-1} \right].
   \label{CD}
\end{equation}
\end{widetext}
Note that both are frequency dependent and have been represented in panels~\ref{Figure3new} (a) and (b) by setting $\tau_p=\tau_n=\tau_0$ for simplicity. Increasing the DC bias reduces the resistance and increases the apparent capacitance due to the $e^{\frac{e V_S}{k_B T}}$ factor in both Eq.~\ref{RD} and \ref{CD} while both collapse in the high frequency limit, $\omega \tau_0 \rightarrow \infty$. 

The corresponding Nyquist plots for the diode current with $m=1$ are shown in Fig.\ref{Figure3new}(c). The negative phase of the impedance confirms the capacitive nature of the drift-diffusion component across the entire frequency spectrum. This characteristic extends to all higher-order modes beyond the fundamental one. 
At the microscopic level, the apparent capacitive response arises because any change in voltage requires the transfer of a certain amount of charge (electrons and holes) to reach a new equilibrium state, which depends on the applied bias~\cite{Bonch-Bruevich1977}. Additionally, as described by Eq.~\ref{jdfin}, the asymptotic behavior for high-frequencies attains the Warburg limit for diffusive transport~\cite{Warburg1901} given by
\begin{equation}
\lim_{\omega \rightarrow \infty} \frac{\Im Z^{(m)}}{\Re Z^{(m)}}=-1.
\end{equation}
This linear trend is illustrated with a dashed line in Fig.~\ref{Figure3new} (c) for the $m=1$ mode.

\begin{figure}[h]
	\includegraphics[width=8.5cm]{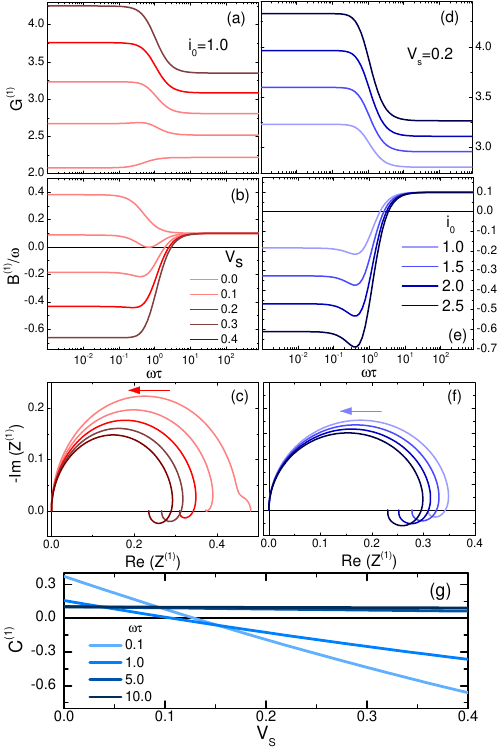}
	\caption{\label{Figure3} First order mode impedance characterization considering the three contributions to Eq.~\ref{jt}. Conductance spectrum (a), susceptance (b), and the corresponding Nyquist maps (c) by varying the DC voltage under fixed illumination condition. Conductance spectrum (d), susceptance (e) and the corresponding Nyquist maps (f) for increasing illumination power under fixed DC voltage. The arrows in panels (c) and (f) indicate the frequency growth direction. (g) Calculated apparent capacitance for the $m=1$ mode as a function of the DC bias for increasing frequency.}
\end{figure}
The results of IS taking into account the contribution of the memristive components introduced in Eq.~\ref{jmfin} are represented in Figs~\ref{Figure3} (a) and (b), which correspond to the Bode plots for the first mode conductance and susceptance, respectively. These plots are obtained by varying the DC bias, $V_S$, and all curves exhibit a transition at $\omega \tau \sim 1$ (on logarithmic scale). To emphasize the high-frequency behavior, the first-mode susceptance is plotted as $B^{(1)}/\omega$. This approach emphasizes the asymptotic trend towards the geometric capacitance, as given by
\begin{equation}
    \lim_{\omega \rightarrow \infty}\frac{B^{(1)}}{\omega}=C_g.
    \label{limite}
\end{equation}
This behavior arises because the reactive contributions to the susceptance in Eq.~\ref{jdfin} grow as $\omega^{1/2}$ for high frequencies, while the displacement contribution grows linearly with frequency. 
The specific parameter choices of $\eta < 0$ and $\alpha= 0.9$ were made to highlight a recurring controversy encountered during solar cell characterization: the interplay between seemingly inductive and capacitive responses observed in impedance spectroscopy. Figure~\ref{Figure3} (b) reveals a noteworthy trend. At higher DC bias values, the first-mode susceptance becomes negative for low frequencies. This behavior can be interpreted as an apparent inductive character in the system. The origin of this negative susceptance lies in Eq.~\ref{jmfin} that contains the reactive contributions to the susceptance. The sign of these contributions is determined by terms proportional to the functions $M(V_S)$ and $N(V_S)$, defined in Eqs.~\ref{M} and~\ref{N}, respectively. In our case, the slight asymmetry introduced by the parameter $\alpha=0.9$ in Eq.~\ref{gen} causes the sign of the generation function (and thus its character) to be dependent on the polarity. This dependence on polarity can lead to a negative susceptance under specific operating conditions, as observed in Fig.~\ref{Figure3}(b). To further illustrate the impact of increasing DC bias on the impedance behavior, Fig.~\ref{Figure3}(c) presents the corresponding Nyquist plots. The arrow indicates the direction of increasing frequency. Many other combinations of intrinsic ($\tau$, $\alpha$, $\eta$, $E_B$) and extrinsic parameters ($\omega$, $V_S$, $V_0$, $T$) not explored here also produce inductive responses.

Thus, in the presence of charge activation, which introduces nonequilibrium carriers into the drift conductance, regardless of whether these carriers are electrons, holes, or ions, an apparent inductive contribution can always be anticipated. However, its effects are confined to the low-frequency range, as indicated by Eq. 15, where all memory contributions vanish at high frequencies, leaving only the geometric capacitance and diffusive channels, both of which exhibit capacitive behavior. The prominence of apparent inductive contributions is enhanced by both amplitude and DC bias. For these contributions to be detectable, they must be at least comparable to other transport mechanisms, a condition influenced by factors such as carrier mobility, effective barrier heights, effective masses, and temperature. This explains the observed transition from an inductive to a capacitive loop in the impedance spectroscopy map shown in Fig.~\ref{Figure3} (c).

Our model also predicts that illumination influences the impedance of the $m=1$ mode of the solar cell. This is not related to the diode density current in Eq.~\ref{jdfin} where the illumination enters just through the term proportional to $g_L$ which doesn't affect the dynamic response (susceptance components). However, Eq.~\ref{jmfin} shows the $g_L$ contribution entering the first-mode conductance without affecting the susceptance. 

Nevertheless, we can prove that illumination might also affect the reactive response. The key lies in the non-equilibrium carrier sources depicted in Figs.~\ref{Figure1}(c) and (d). Illumination fills these states, reducing the effective barriers' values and this directly impacts the term $i_0$ in Eq.~\ref{gen}. The factor $i_0$ is proportional to $\exp(-E_B/k_b T)$ and influences the weighting of functions $M$ and $N$ within the memristive current term of Eq.~\ref{jmfin}. Thus, illumination indirectly affects the dynamic response (susceptance) through its influence on the effective barrier heights. Figures~\ref{Figure3}(d), (e), and (f) further illustrate this point. These plots map the Bode/Nyquist impedance response as $i_0$ increases, demonstrating an enhanced apparent inductive character with illumination. Similar enhancement of apparent capacitive trends could also be expected by using the same arguments.

\begin{figure}[h]
	\includegraphics[width=8.5cm]{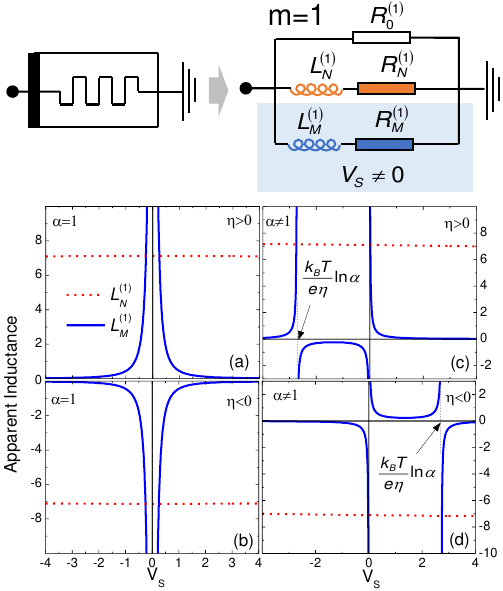}
	\caption{\label{Figure4} Upper panel: apparent circuit representation of the first order mode of a memristive channel. Apparent inductance values for: symmetric charge transfer, with $\alpha=1$, for (a) $\eta>0$ and (b) $\eta<0$; asymmetric charge transfer, with $\alpha\neq 1$, for (c) $\eta>0$ and (d) $\eta<0$.}
\end{figure}

\begin{figure}[h]
	\includegraphics[width=8.5cm]{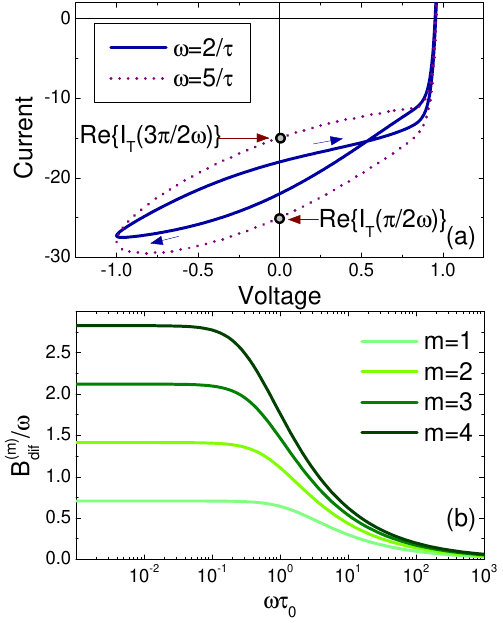}
	\caption{\label{Figure5} (a) Cyclic voltammetry characterization of the solar cell by increasing frequency, highlighting the widening of the short-circuit current splitting. (b) First four orders contributions to the diffusive susceptance.}
\end{figure}

The transition from a capacitive-like to an inductive response in the device impedance is often described in the literature as the emergence of negative capacitance~\cite{Ebadi2019,MoraSero2006,Jonscher1986,Joshi2020,Ershov1998}. This correlation arises when defining the apparent capacitance per mode as $C^{(m)}=\Im(1/Z^{(m)})\omega^{-1}$. As shown in panel 4(g) form $m=1$, this value transitions from positive to negative at lower frequencies. However, we argue that the most accurate way to describe these trends in terms of apparent circuits is by correlating this tuning to a fixed (frequency-independent) apparent inductive element.

Unlike the diode contribution in Eq.~\ref{jdfin}, that does not allow for the segmentation of the dynamic response in terms of apparent elementary circuit components independent on frequency, the memristive components in Eq.~\ref{jmfin} allow for such segmentation. This is illustrated in Figure~\ref{Figure4} (top panel) for the m=1 mode. Here, the second line of Eq.~\ref{jmfin} is represented by an apparent circuit with elements corresponding to specific terms in the equation: $R_0^{(1)}=[A \gamma \left( N_0 + g_L \tau + g_0 \tau \right)+A N/2]^{-1}$, $L_N^{(1)}=8 \tau/(A\cdot N)$, $R_N^{(1)}=L_N^{(1)}/(2 \tau)$, $L_M^{(1)}=V_0 \tau/(2 M \cdot A V_S)$, and $R_M^{(1)}=L_M^{(1)}/\tau$. Note that the lower branch ($L_M^{(1)}$) is not conductive at zero DC bias ($V_S=0$). Additionally, the apparent inductances and resistances can be positive or negative depending on the values of $N$ and $M$. This dependence is shown in Figures~\ref{Figure4} (a) and (b) for a symmetric generation function $\alpha=1$. These figures depict the behavior for pure generation ($\eta>0$) and pure trapping ($\eta<0$) scenarios. At low voltages, conduction primarily occurs through the $L_N^{(1)}$ branch, while at higher biases the $L_M^{(1)}$ branch dominates. The asymmetric case ($\alpha \neq 1$), displayed in Figs.~\ref{Figure4} (c) and (d) is more complex. Here, the model predicts the possibility of tuning the character (positive or negative inductances) of the prevailing conductive branch. Additionally, the model foresees a singular point at $V_S=\frac{k_B T}{e \eta}\ln \alpha$, where the character undergoes a second inversion. These points correspond to the extrema of the generation function ($d g/dV=0$) where the dynamic conductance component of the first mode, proportional to $M$, vanishes.

To fully capture the intricacies of these systems, we should consider concurrent memory channels with diverse characteristics, such as varying relaxation times and non-equilibrium carrier transfer behavior (combining trapping or activating nature). These complexities were symbolically represented in Fig.~\ref{Figure1} (a) as $j_{M_i}$ ($i=$1,2...) and are beyond the scope of the present discussion. Furthermore, other mechanisms can contribute to apparent inductive effects in the transport response of diodes under cyclic biasing. A notable example is the generation of additional carriers via impact ionization within avalanche diodes at high electric fields, as explored in Ref.~\citenum{Pikus1965}.

Figure~\ref{Figure5} illustrates the evolution of hysteresis as the cycling frequency increases. At lower frequencies, an apparent inductive hysteresis at positive bias, driven by nonequilibrium charge activation under these conditions, coexists with a capacitive transport component that reflects both geometric and diffusive contributions. However, as the frequency increases, the influence of charge activation diminishes, allowing the capacitive behavior to dominate.

Our theoretical framework also allows for qualitative and quantitative characterization of the apparent capacitive effects observed under short-circuit conditions ($V=0$). This is evident in the splitting of the total current for down and up-voltage sweeps, denoted as $\Re{I_T(\pi/2 \omega)}$ and $\Re{I_T(3\pi/2 \omega)}$ in Fig.~\ref{Figure5} (a) (assuming $t=0$ at the beginning of each voltage cycle). According to Eq.~\ref{IT}, the short-circuit current splitting arises solely from the contribution of the susceptance of odd modes
\begin{eqnarray}
    \Delta I_{SC}(\omega) &=& \Re{I_T\left(\frac{3\pi}{2 \omega}\right)}-\Re{I_T\left(\frac{\pi}{2 \omega}\right)} \nonumber \\
    &=& 2 V_0 \sum_{k=1}^{\infty} (-1)^{k+1} B^{(2k-1)}(\omega),
    \label{delti}
\end{eqnarray}
where $B^{(m)}(\omega)=B^{(m)}_{dif}(\omega) + C_g \omega \cdot \delta_{m,1}$, with $B^{(m)}_{dif}$ representing the diffusive contribution to the susceptance of the $m$-th mode, that by following Eq.~\ref{jdfin} and assuming $\tau_p=\tau_n=\tau_0$, can be expressed as
\begin{equation}
    B^{(m)}_{dif}(\omega)= \frac{A}{\sqrt{2}V_0}\frac{j_S}{m!}\left(\frac{eV_0 }{k_B T}\right)^m \sqrt{\sqrt{1+(m\omega \tau_0)^2}-1}.
\end{equation}
Note that irrespective of the hysteresis complexity produced by the memristive component as displayed in Fig.~\ref{Figure5} (a), $j_M$ cannot contribute to the short-circuit current splitting ($\Delta I_{SC}$) due to its definition in Eq.~\ref{jm}. This is correctly captured in the third-order approximation presented in Eq.~\ref{jmfin}, where the susceptance components of the first and third-order modes are identical. Consequently, these terms cancel each other out when applying the definition of $\Delta I_{SC}$ in Eq.~\ref{delti}.

Figure~\ref{Figure5} (b) displays the diffusive susceptance normalized by frequency ($B^{(m)}_{dif}/\omega$). If the contribution of modes $m > 1$ can be neglected (e.g., at low enough amplitudes), the short-circuit current splitting can be easily used to assess the relative impact of geometric and apparent diffusive capacitances. This is achieved through the following relationship obtained by considering just the first term in the sum of Eq.~\ref{delti}
\begin{equation}
\frac{\Delta I_{SC}}{2 V_0 \omega}\simeq C_g+\frac{B^{(1)}_{dif}(\omega)}{\omega},
\label{delt}
\end{equation}
where a correlation with the spectroscopic results from Eq.~\ref{limite} becomes evident.

A deviation from a linear increase of $\Delta I_{SC}$ with frequency signifies a non-negligible contribution from the diffusive channels to the apparent capacitance. This can be understood by examining Eq.~\ref{delt} (or Eq.~\ref{jdfin} for the definition of $B^{(1)}_{dif}$). Here, the diffusive contribution to the susceptance, $B^{(3)}_{dif}$, scales as the square root of frequency (i.e., $B^{(m)}_{dif}= O(\omega^{1/2})$) and can be singled out by substracting the constant finite limit, $C_g$, obtained from Eq.~\ref{delt} at high frequencies. Note that, according to Fig.~\ref{Figure5} (b), the inflection point of $B^{(m)}_{dif}(\omega)/\omega$ can be used as a reference of the recombination time $\tau_0$. Specifically, the condition $d^2/d \omega^2 [B^{(m)}_{dif}(\omega)/\omega]=0$ occurs at $\omega \tau_0=1/(6 m) \sqrt{6+6\sqrt{33}}$, which corresponds to $\omega \tau_0 \simeq 1.0602$ for $m=1$.

In summary, our work presents a unified theoretical framework for analyzing both cyclic voltammetry and impedance spectroscopy of solar cells. This approach incorporates diffusive transport and non-equilibrium carrier behavior to explain seemingly anomalous inductive or capacitive responses. We have derived comprehensive analytical expressions for current-voltage relationships and developed a multimode spectral analysis of the complex admittance. Furthermore, the model organically integrates the influence of DC biasing and illumination, predicting how these external factors significantly modulate the device's impedance. This approach, with its focus on non-equilibrium effects, extends beyond the specific cases of perovskites and memristors, offering a versatile framework containing elements, which are applicable to a wide range of photovoltaic technologies and device architectures.
We hope this unified framework can pave the way for a deeper understanding and optimization of solar cell performance.

\textbf{Acknowledgments} This study was financed in part by the Coordenação de Aperfeiçoamento de Pessoal de Nível Superior - Brazil (CAPES) and the Conselho Nacional de Desenvolvimento Científico e Tecnológico - Brazil (CNPq) Proj. 311536/2022-0.

\bibliography{mainResub.bbl}


\end{document}